\documentclass[11pt]{article} 
\usepackage{hyperref} 

\begin{document} 
\title{Inertially-induced secondary flow in microchannels \\ Fluid Dynamics Videos} 
\author{Hamed Amini, Mahdokht Masaeli, Dino Di Carlo \\ 
\\\vspace{6pt} Department of Bioengineering\\
University of California, Los Angeles, CA 90095, USA}

\maketitle
\begin{abstract}
\noindent We report a novel technique to passively create strong secondary flows at moderate to high flow rates in microchannels, accurately control them and finally, due to their deterministic nature, program them into microfluidic platforms. Based on the flow conditions and due to the presence of the pillars in the channel, the flow streamlines will lose their fore-aft symmetry. As a result of this broken symmetry the fluid is pushed away from the pillar at the center of the channel (i.e. central z-plane). As the flow needs to maintain conservation of mass, the fluid will laterally travel in the opposite direction near the top and bottom walls. Therefore, a NET secondary flow will be created in the channel cross-section which is depicted in this video.\\
\noindent The main platform is a simple straight channel with posts (i.e. cylindrical pillars Ð although other pillar cross-sections should also function) placed along the channel. Channel measures were 200$\mu$m$\times$50$\mu$m, with pillars of 100$\mu$m in diameter. Positioning the pillars in different locations within the cross-section of the channel will result in induction of different secondary flow patterns, which can be carefully engineered. The longitudinal spacing of the pillars is another design parameter (600$\mu$m spacing was used for this video). The device works over a wide range (moderate to high) flow rates. We used 150$\mu$L/min in this experiment. The device has 3 inlets where a dye stream is co-flowed between two water streams. In this video, one can see the effect of the net secondary flow created by inertia in the microchannel by visualizing the cross-section of the fluorescently labeled stream. Confocal images are sequentially taken at the inlet and after 48 consecutive pillars.
\end{abstract}

\end{document}